\documentclass{jfm}
   \usepackage{graphicx}
   \usepackage{amsmath,amssymb}
   \usepackage{natbib}
   \usepackage{latexsym}
   \usepackage{url}
   \usepackage{bm}
   \usepackage{color}
   \usepackage{epstopdf}
   \usepackage[margin=0cm,justification=justified]{caption}
   \graphicspath{{../../figures/}}

\title{Lifetime of localized turbulence in transitional channel flow}
\author{Yuze Wu
 \and Baofang Song \corresp{\email{baofang.song@pku.edu.cn}}}

\affiliation{State Key Laboratory for Turbulence and Complex Systems, and HEDPS, Center for Applied Physics and Technology, School of Mechanics and Engineering Science, Peking University, Beijing 100871, China}

\begin{document}
\maketitle
\begin{abstract}
The lifetime of localized turbulence in transitional channel flow, the so-called turbulent stripes or bands, as a function of the Reynolds number ($Re$) is reported here.
Recent studies showed that although the mean lifetime of a turbulent band increases with the band length, it appears to saturate at a certain critical length, which is smaller than the fully developed band length. However, this observation was based on a single Reynolds number only. Here, we first confirmed this saturation by considering higher Reynolds numbers. This property enables us to measure the characteristic lifetime of a turbulent band without having to resolve it at its full length, which greatly reduces the computational cost and makes statistical studies by direct numerical simulation (DNS) feasible near the lower end of the transitional-$Re$ regime. Subsequently, by extensive DNS, we measured the lifetime up to $Re=670$, and the  data cover more than two orders of magnitude, ranging from $O(10^2)$ to nearly $O(10^5)$ and revealing a sharp but smooth growth with increasing $Re$. No sudden divergence of the lifetime was observed in the $Re$ range investigated, in which controversy existed over whether or not a turbulent band would become sustained. Despite significant differences between the structure and dynamics of channel bands and pipe puffs, our data concretely support a unified transition scenario in the two flows.

\end{abstract}

\indent 

\section{Introduction}\label{sec:introduction}
Turbulent puffs in pipe flow are known to be transient, undergoing memoryless decay, and have a characteristic lifetime that grows super-exponentially with the Reynolds number \citep{Peixinho2006, Hof2006, Willis2007, Hof2008, Avila2010}. The balance between this transient mechanism with the proliferation mechanism determines the critical Reynolds number for the onset of sustained turbulence \citep{Avila2011,Avila2023}. It has been expected that all other canonical shear flows share this unified transition scenario. However, in channel flow, the temporal characteristics of a fully localized turbulent band are much less clear, with even conflicting views on whether a fully developed band would remain transient or become sustained above a critical Reynolds number \citep{Tao2018, Kanazawa2018, Paranjape2019, Shimizu2019, Xu2022, Mukund2025, Wu2025}.

Different from pipe puffs, which are highly confined by the pipe geometry and behave quasi-one-dimensionally on large scale, channel bands, being less confined by the solid walls, can extend in both streamwise and spanwise directions and therefore have distinct complexities. More specifically, localized turbulence in channel flow organizes itself into a banded structure obliquely oriented about the streamwise direction \citep{Tsukahara2005, Tuckerman2014}. 
In a sufficiently large channel, a turbulent band develops into a fully localized structure (i.e. finite length) with an elongated bulk part, an upstream end, and a downstream end
\citep{Xiong2015, Tao2018, Kanazawa2018, Paranjape2019, Shimizu2019, Xiao2020, Xiao2020b, Song2020, Kashyap2020, Marensi2023, Mukund2025, Wu2025}. But the dynamics at the two ends of a band are asymmetric: At the downstream end (referred to as {\it head} hereafter), turbulence is generated continuously and quasi-periodically via a seemingly deterministic instability mechanism, whereas turbulence simply decay or the tail part chips intermittently at the upstream end (referred to as {\it tail} hereafter) \citep{Kanazawa2018, Shimizu2019, Paranjape2019, Xiao2020, Xiao2020b, Song2020, Marensi2023, Mukund2025}. The bulk part of a band exhibits nearly wave-like alternating high- and low-speed streaks and vortices \citep{Kanazawa2018, Song2020, Xiao2020b, Marensi2023}. The more regular structure and deterministic growth mechanism render it distinct from pipe puffs and Couette-flow stripes \citep{Marensi2023}. 

\begin{figure}
    \centering
    \includegraphics[width=0.75\linewidth]{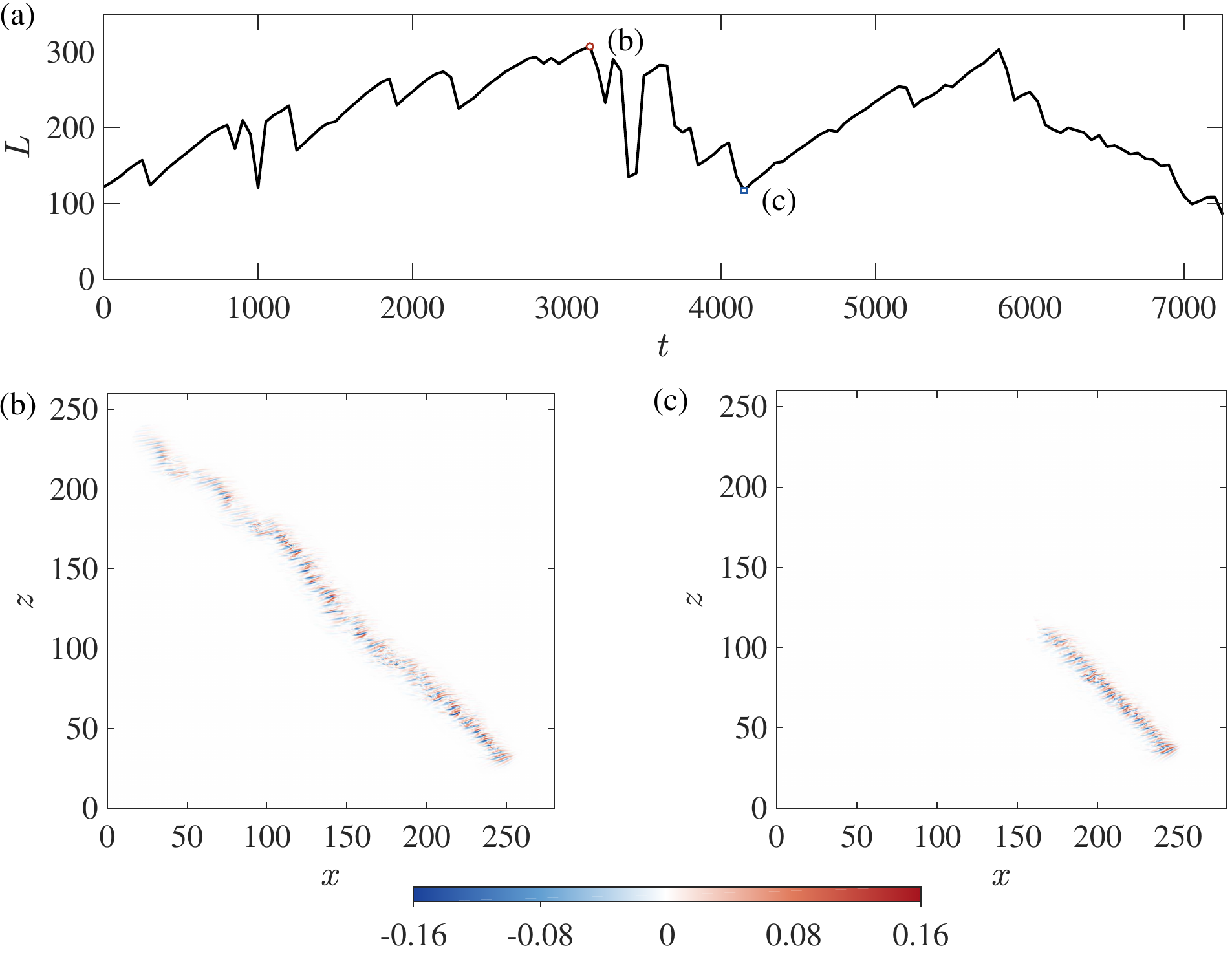}
    \caption{\label{fig:band_length} The length of a fully developed turbulent band at $Re=640$. (a) Time series of the band length, which show large fluctuations of the band length, between $L\simeq 100-300$ in unit of half channel height, due to the decay or chipping of turbulence at the tail. (b) The visualization of the band in a cross-section parallel to the channel walls at the instant of maximum band length, labeled as (b) in panel (a). In the figure, $x$ denotes the streamwise coordinate and $z$ denotes the spanwise coordinate, and the flow is from left to right. Contours of the wall normal velocity fluctuations in the cross-section are plotted. (c) The same plot as in panel (b) for the band at the instant of its minimum length, labeled as (c) in panel (a).}
\end{figure}

The main difficulty in studying the temporal characteristics of turbulent bands lies in their large length and time scales. A fully developed band was shown to have a length of about O(100) channel heights (see figure \ref{fig:band_length}(a) for an example at $Re=640$), which increases with $Re$. For example, the time-averaged length at $Re=660$ was reported to be about 150 channel heights, and it increases to about 220 channel heights at $Re=720$. Furthermore, the strongly fluctuating instantaneous length can be considerably larger than the time average \citep{Kanazawa2018} (see figure \ref{fig:band_length}). Besides, the band can evolve more than $10^4$ convective time units without a sign of decay at $Re\gtrsim 660$ \citep{Tao2018, Kanazawa2018, Paranjape2019,  Xiao2020}.
These two factors make it difficult to perform either experimental or DNS studies on the temporal characteristics of a channel band at its full length, especially statistical studies that require large samples of realizations and sufficiently long observation times. The controversies in former studies \citep{Tao2018, Kanazawa2018, Paranjape2019, Shimizu2019, Kashyap2020, Xu2022, Mukund2025} over whether a turbulent band remains transient or becomes sustained above a critical $Re$ and the value of this $Re$ could be due to different observation times and lack of sufficient statistics.

To manage statistical studies with large samples and long observation times, \citet{Xu2022} used a damping technique to cut the tail part of a turbulent band so as to restrict the band length. This consideration was based on the observation that the 
turbulence-generation mechanism at the head is crucial for the growth and self-sustenance of the entire band, without which a band will quickly decay at $Re\lesssim 1000$ \citep{Shimizu2019, Song2020, Wu2022,  Kohyama2022, Marensi2023, Mukund2025}. The dynamics at the head and in the close neighborhood was shown to determine the tilt angle of the band also \citep{Zhang2023}. Therefore, it is essential to keep the head part of a turbulent band, whereas the tail part should be less important because turbulence just decays there. \cite{Xu2022} observed that such a moderately constrained band exhibits a memoryless transient behavior, similar to pipe puffs. Nevertheless, the lifetime increases sharply with the band length, indicating that the bulk part certainly influences the sustenance of a band, and even the tail part might have certain influence also. This raises another question: If the lifetime sensitively depends on the band length, will a fully developed turbulent band have a characteristic lifetime given that its length fluctuates significantly over time due to the decay and intermittent chipping of the tail part? 

Considering that the band lengths studied in \citet{Xu2022} were much shorter than the fully developed ones and therefore were not sufficient to conclude how a fully developed band would behave, \citet{Wu2025} further pushed to significantly larger band lengths at $Re=640$. The domain size they used, though still not large enough to fill in a fully developed band, allows the band to temporarily vary its length due to the intermittent chipping at the tail. By extensive DNS, the authors reported memoryless transient behavior of the band and  saturation of the mean lifetime as the band length is increased. This saturation is particularly important because it suggests that a band, even with varying length, may have a characteristic lifetime once its length grows to above a critical value. However, \citet{Wu2025} only considered a single $Re$ and little is known about whether or not this saturation of lifetime is a general characteristic in the transitional regime.

In this paper, by extensive DNS, first, we extend to higher Reynolds numbers to verify the saturation of the lifetime, and thus to support that a turbulent band should have a characteristic lifetime. Second, we measure the saturated lifetime as a function of $Re$ to clarify whether a turbulent band remains transient in the $Re$ range where controversies existed.


\section{Methods}
\label{sec:methods}
We use the numerical method of \citet{Xu2022} and \citet{Wu2025}. For spatial discretization, 64 Chebyshev grid points and finite-difference method with a 9-point stencil are used in the wall-normal direction ($y$). Periodic boundary conditions and Fourier spectral methods are used in the streamwise ($x$) and spanwise ($z$) directions with grid sizes $h/\Delta x\simeq 5.3$ and $h/\Delta z\simeq 7$ before dealiasing in these two directions, respectively. The resolution is close to those used for comparable Reynolds numbers \citep{Tao2018, Kanazawa2018, Shimizu2019, Kashyap2020, Marensi2023}. A second-order time-stepping scheme is used with the time-step size fixed at $\Delta t=0.025$, and the viscous term is treated implicitly while the nonlinear term is treated explicitly \citep{Hugues1998}. 
The mass flux is fixed by dynamically adjusting the driving pressure gradient in the $x$-direction. More details about the numerical method can be found in our earlier work \citep{Xiao2020, Xu2022, Wu2025}. 

\begin{figure}
    \centering
    \includegraphics[width=0.98\linewidth]{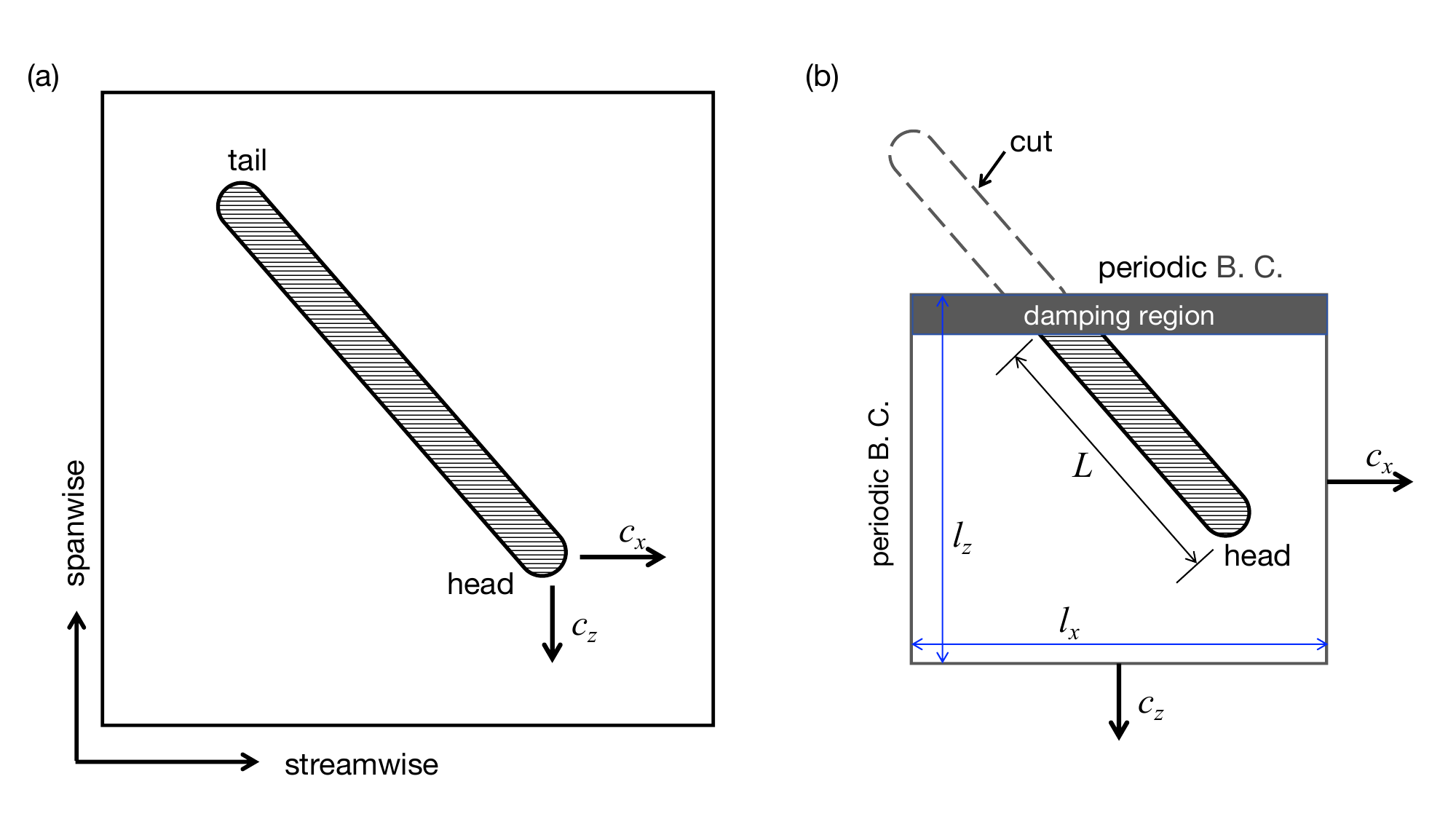}
    \caption{\label{fig:sketch} Sketch of the simulation technique. (a) A fully developed band in the laboratory frame of reference, in which the head of a turbulent band travels with a streamwise speed $c_x$ and spanwise speed $c_z$. (b) The moving frame of reference with a damping region used in our simulations, in which the head of a band is fixed with a constant distance to the damping region. The length of the band $L$ can be controlled by properly choosing box sizes ($l_x$ and $l_z$) and setting the position and size of the damping region. Note that the band is tilted about the streamwise direction by an angle of approximately $40^\circ$ in the considered $Re$ regime. This sketch is adapted from figure 1 of \citet{Wu2025}.}
\end{figure}

The computation technique used for tracking the head of a turbulent band and controlling the band length as well as the underlying consideration were detailed in \citet{Wu2025}. For the ease of discussion, we still show a schematic diagram of the computation setting in figure \ref{fig:sketch}. A narrow damping region with a spanwise width of $10$ is used to cut the tail part of a band. We realize $L\simeq 85$ by using $(l_x, l_z)=(100,100)$, $L\simeq 110$ by using $(l_x, l_z)=(120,110)$, and $L\simeq 140$ by using $(l_x, l_z)=(140,130)$. 

This computation technique enables us to monitor the traveling speed of the head in runtime, which was shown to be a good diagnosis for the decay of a turbulent band \citep{Wu2025}. The normal spanwise speed is $|c_z|\simeq 0.1$ and fluctuates around slightly \citep{Tao2018, Paranjape2019, Shimizu2019, Xiao2020, Xiao2020b, Song2020}, but it will significantly decrease when the turbulence starts to decay. The criterion for a decay event is set to be $|c_z|<0.05$ in this study because turbulence never recovers once $|c_z|$ drops below 0.05 in our simulations.

\section{Results}\label{sec:results}

\begin{figure}
    \centering
        \includegraphics[width=0.99\linewidth]{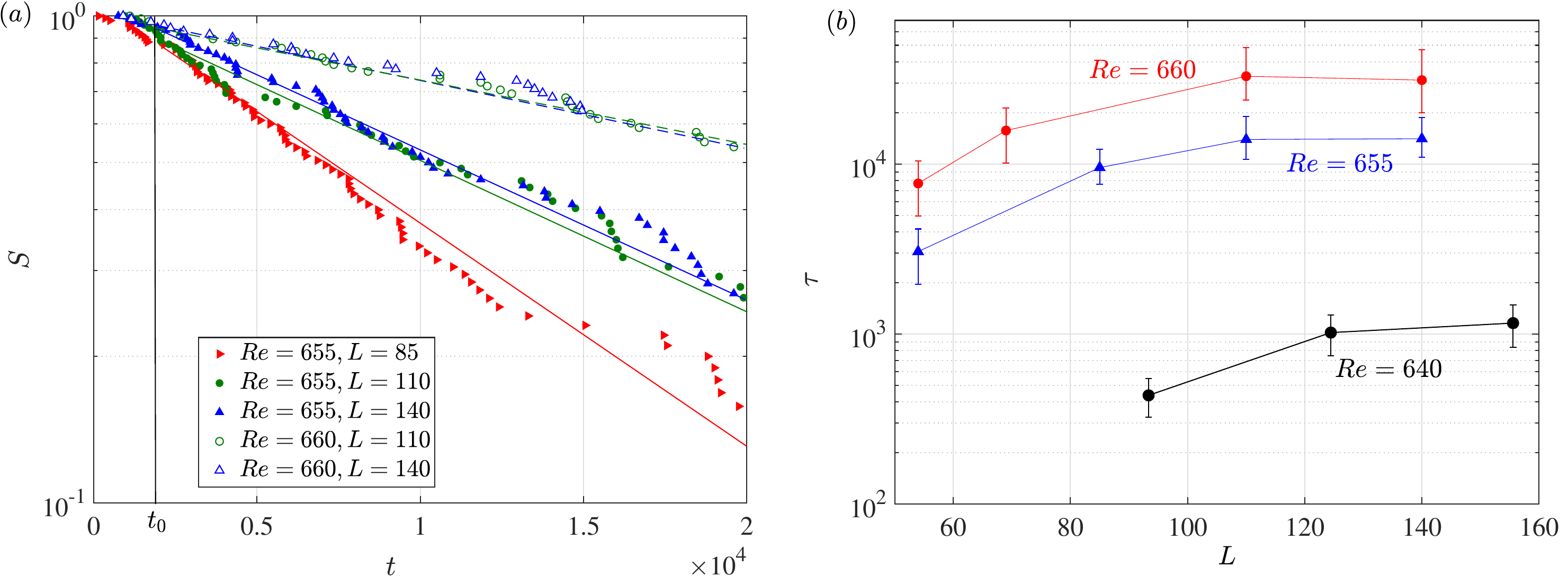}
    \caption{\label{fig:saturation} Saturation of the lifetime with respect to the band length $L$. The survival rate $S$ as a function of time $t$ for $Re=655$ and 660 (a) and the lifetime as a function of the band length $L$ (b). In panel (a), lines are the respective exponential fits $S(t_0)\exp{(-(t-t_0)/\tau)}$. The vertical line shows the time $t_0=2000$. For comparison, the data for $Re=640$ from \citet{Wu2025}, the data for $Re=655$ with $L=54$, and the data for $Re=660$ with $L=54$ and 69 from \citet{Xu2022} are also included in panel (b). The errorbars show the 95\% confidence interval of the estimation of $\tau$.}
\end{figure}

Saturation of the lifetime about the band length was shown for a single Reynolds number of $Re=640$ in \citet{Wu2025}. Here, we first show further verification of this property at higher $Re$'s. For $Re=655$, the band lengths 
$L\approx 85,~110$, and 140 are considered. For each $L$, simulations are repeated many times with different initial conditions, which are taken at different time instances from long-lived turbulence at the same $Re$ but slightly perturbed. Due to the chaotic nature of turbulent dynamics, these trajectories will quickly deviate from the source trajectory and decorrelate. The full sample sizes are $N=95, 72$, and 78 for the three lengths, respectively. Figure \ref{fig:saturation}(a) shows the survival rate (probability) $S(t)=P\{\mathrm{survival~time}\ge t\}$ as a function of time. All three cases show approximately exponential tails over the observation time (up to $2\times 10^4$ convective time units), and therefore, the lifetime can be well described by an exponential distribution. A major difference from the analysis of \citet{Wu2025} is that, due to the large survival times of the turbulence at $Re=655$, we cannot afford to simulate until all turbulent bands have decayed. Instead, we set a truncation point of the observation time and use the maximum-likelihood estimation with censored data as described in \citet{Lawless2003} and \citet{Avila2010}. That is, the lifetime is estimated as

\begin{equation}
\label{equ:lifetime}
\tau=\frac{1}{r}\left[\sum_{i=1}^r(t_{d,i}-t_0)+(N_s-r)(T-t_0)\right],
\end{equation}
where $N_s$ is the size of the subsample that contains the cases where turbulence lives longer than a threshold $t_0$, which is meant to remove the effects of the initial adjustment of the flow and exclude the cases where turbulence starts to decay nearly immediately after the initiation. This subsample will be used for the actual lifetime statistics. In equation (\ref{equ:lifetime}), $r$ is the number of cases in which turbulence decays within the observation time, $t_{d,i}$ is the time instance of the $i^{\mathrm{th}}$ decay event, $T$ is the truncation time, and $N_s-r$ gives the number of cases in which turbulence has not decayed until the truncation time, see figure \ref{fig:saturation}(a) in which $T=20\mathrm{,}000$. 

When taking $t_0=2000$ (larger $t_0$ gives no significant difference in the estimated $\tau$ except for smaller $N_s$ and consequently larger error bars), the sub-sample sizes actually used for the lifetime statistics are $N_s=83,~65$, and 73, respectively. The resulting mean lifetime and the errorbars are shown in figure \ref{fig:saturation}(b). A saturation trend is indeed observed as $L$ increases, similar to the $Re=640$ case reported in \citet{Wu2025}. This observation supports that there is indeed a critical band length above which the lifetime does not change anymore, giving a characteristic lifetime of a turbulent band even if the band length varies with time above the critical value.

Next, we show the lifetime measurement at $Re=660$, for which we only considered the length $L=110$ and 140, and the full sample sizes are both $N=78$ and 72, respectively. The purpose is to see whether the lifetime has also saturated in this length range. Truncation time is set as $T=20000$ for the $L=110$ case while $T=15000$ for the $L=140$ case considering the much larger computation cost. For better showing the trend, we collected the data for $L=54$ and $69$ from \citet{Xu2022}, which were based on only 30 realizations but without a truncation time, i.e. turbulence decayed in all realizations and the mean lifetime was estimated by the sample mean directly. The results are also included in figure \ref{fig:saturation}(b). Compared to $Re=655$, decay event is observed in a significantly smaller fraction of realizations, which is only about 50\% within $T$ and results in smaller $N_s$'s and larger error bars. It can be seen that the survival rate curves for $L=110$ and 140 approximately collapse onto each other and the estimated lifetime are statistically equal (i.e. indistinguishable within the statistical uncertainty). Both values are significantly larger than the lifetime with the other two smaller $L$'s. Therefore, the data suggest that $\tau$ also tend to saturate as the band length increases. From the trend shown in figure \ref{fig:saturation}(b), the critical band length for saturation appears to be some value around $L\simeq 110-120$ in the considered $Re$ range. 

\begin{figure}
    \centering
        \includegraphics[width=0.99\linewidth]{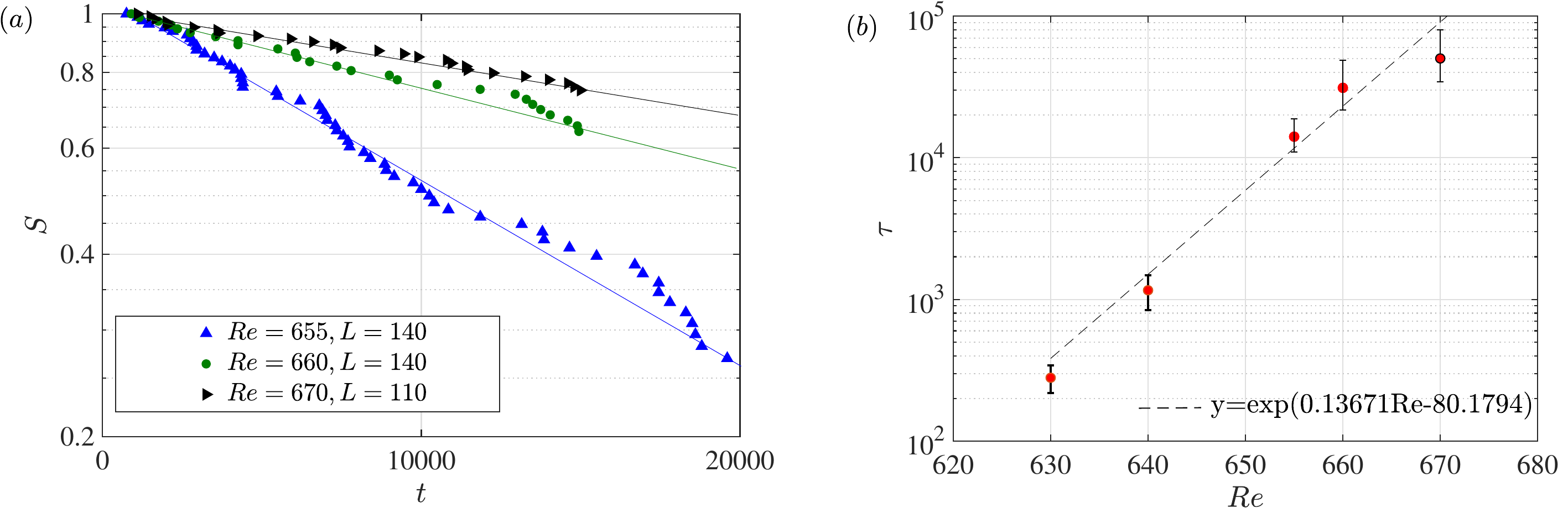}
    \caption{\label{fig:lifetime} Panel (a) shows the survival rate $S$ for $Re=670$ with $L=110$. Data for $Re=655$ and 660 with $L=140$ are also included for showing the trend as $Re$ increases. Panel (b) shows the lifetime $\tau$ as a function of $Re$, with the data for $Re=630$ and 640 from \citet{Wu2025} also included. The dashed line in panel (b) shows an exponential fit to the lifetime data.}
\end{figure}

As $Re$ increases further, similar DNS studies will become prohibitively expensive and we cannot afford such a lifetime-saturation study. Nevertheless, noticing that the length at the saturation is around $L\simeq 110$ for $Re=655$ and 660, we will assume that the saturation length is the same at close Reynolds numbers and measure the lifetime at $Re=670$ by using $L=110$. Consequently, one should note that the measured lifetime might have been underestimated. Besides, considering the large lifetime, we have to use a shorter truncation time of $T=15000$ to reduce the cost further. With larger lifetime, there will be fewer decay events within the truncation time $T$. To assure that there are sufficient decay events for the statistics so that the uncertainty is practically meaningful, the sample size of $N=99$ for $Re=670$ is used, which gives 22 decay events between $t_0$ and $T$ (i.e. $r=22$ in equation \ref{equ:lifetime}). The $S(t)$ is shown in figure \ref{fig:lifetime}(a), which can be fitted as an exponential function also.

With the data obtained here and collected from our previous work \citep{Wu2025}, covering more than two orders of magnitude in the lifetime, we present $\tau=\tau(Re)$ in figure \ref{fig:lifetime}(b). Clearly, the lifetime remains finite at all $Re$'s and reaches $O(10^5)$ at $Re=670$ already, confirming the transient nature of localized turbulence in channel flow at least in this Reynolds number regime. An exponential fit can roughly describe the trend of the data.

\section{Discussion and conclusion}

Through extensive computer simulations at multiple transitional Reynolds numbers, we show evidences for the existence of a critical band length above which the lifetime saturates with respect to the band length, indicating a characteristic mean lifetime for a fully developed band in the transitional regime. In the studied $Re$ range, this critical length is found to be $L\simeq 110-120$.
Taking this saturated lifetime as the characteristic lifetime, we obtained the relationship $\tau=\tau(Re)$ for localized turbulence in transitional channel flow.

The results show a rapid but smooth increase in the lifetime by more than two orders of magnitude over a rather narrow Reynolds number range of $Re=630-670$. The increasingly large but finite lifetime clarifies the controversies over whether or not there exists a sudden transition to a sustained turbulent band in this $Re$ regime \citep{Kanazawa2018, Tao2018, Paranjape2019, Shimizu2019, Kashyap2020, Xu2022, Wu2025}. The results support the same transition scenario as known for pipe flow, that is, an individual localized turbulence in channel flow also remains transient as $Re$ increases. Although this scenario has been reported already for strongly confined channel bands without the head and tail in narrow computational domains \citep{Gome2022}, our results, for the first time to our knowledge, show concrete evidences for the transient nature of a fully developed turbulent band. What follows is that the sustained turbulent state in channel should be reached by either the proliferation rate of a band outweighing its decay rate or a delocalizing growth mechanism of a band outweighing its decay like the puff-slug transition in pipe flow \citep{Avila2011, Barkley2015}, depending on which scenario takes place earlier as $Re$ increases. Former computations in large channel domains \citep{Shimizu2019, Kashyap2020, Kohyama2022} and experiments \citep{Paranjape2019} appeared to suggest the former. 

Although the memoryless transient characteristic is confirmed, the precise scaling of the $\tau=\tau (Re)$ cannot be concluded in such a narrow $Re$ range. Although an exponential scaling can roughly describe the data as shown in figure \ref{fig:lifetime}(b), given the few data points and the statistical uncertainties, one can fit something else also. Due to the slow convergence of the uncertainty with respect to the sample size, much larger samples and also a larger $Re$ range are required for obtaining a conclusive scaling law. However, given that the lifetime is almost O($10^5$) at $Re=670$, resolving this scaling is out of the reach of our computation power. Nevertheless, using the fitted exponential scaling as a guide, one can expect a lifetime of $\tau>1\times 10^5$ at $Re\gtrsim 680$, which is consistent with the observation of \citet{Shimizu2019} based on long-time simulations ($\approx 1.5\times 10^5$ time units) that turbulence would decay below $Re\simeq 700$ but appeared to be sustained otherwise.
  
A strict determination of the critical point for sustained turbulence is still challenging, which requires measuring the timescale of turbulence proliferation as well, as done for pipe flow \citep{Avila2011} and strongly confined channel turbulence \citep{Gome2020}. Although our computation technique enables to measure the lifetime, it would fail for the study of longitudinal splitting, which occurs necessarily at the tail part of a band \citep{Paranjape2019, Shimizu2019, Mukund2025}. Therefore, the tail cannot be simply damped and a band at its full length is probably needed for studying the longitudinal splitting, which will be extremely expensive for a similar statistical study. The additional proliferation route via transverse branching \citep{Paranjape2019, Shimizu2019}, which may not be separated from the longitudinal splitting, further complicates the problem. Therefore, so far, the critical point of channel flow still cannot be neatly determined by balancing the decay and proliferation processes of a single localized turbulence. Another related question is if channel-flow transition can be categorized as a directed percolation (DP) phase transition  \citep{Hinrichsen2010} like the transition of Couette flow \citep{Lemoult2016, Klotz2022, Hof2022} and pipe flow \citep{Lemoult2024}. Although this scenario was claimed for channel flow in some studies \citep{Sano2016, Shimizu2019, Kohyama2022}, doubts remain because the studies were not able to approach sufficiently close to the critical point (still unknown yet) and the thermodynamic limit. In fact, a single type of phase transition  even may not be sufficient to describe the transition process, which was proposed to be multiple-staged and more complex than other canonical shear flows \citep{Shimizu2019, Manneville2020}. In a word, channel-flow transition is highly nontrivial and many aspects of this process remain to be clarified.


\label{sec:conclusion}

\section{Acknowledgement}
The authors acknowledge financial support from the National Natural Science Foundation of China under grant number 12322209, 12588201, and 12272264. The work is also supported by The Fundamental Research
Funds for the Central Universities, Peking University. The simulations were performed on supercomputers provided by Beijing Paratera Technology Co., Ltd. and SuperComputing
Network.

\section*{Declaration of interests}
The authors declare no conflicting interests.

\bibliographystyle{jfm}
\bibliography{references}

\begin{thebibliography}{37}
\expandafter\ifx\csname natexlab\endcsname\relax\def\natexlab#1{#1}\fi
\def\au#1{#1} \def\ed#1{#1} \def\yr#1{#1}\def\at#1{#1}\def\jt#1{\textit{#1}}
  \def\bt#1{#1}\def\bvol#1{\textbf{#1}} \def\vol#1{#1} \def\pg#1{#1}
  \def\publ#1{#1}\def\arxiv#1{#1}\def\org#1{#1}\def\st#1{\textit{#1}}

\bibitem[Avila {\em et~al.\/}(2011)Avila, Moxey, De~Lozar, Avila, Barkley \&
  Hof]{Avila2011}
{\sc \au{Avila, K.}, \au{Moxey, D.}, \au{De~Lozar, A.}, \au{Avila, M.},
  \au{Barkley, D.} \& \au{Hof, B.}} \yr{2011}  \at{The onset of turbulence in
  pipe flow}.  \jt{Science}  \bvol{333},  \pg{192--196}.

\bibitem[Avila {\em et~al.\/}(2023)Avila, Barkley \& Hof]{Avila2023}
{\sc \au{Avila, M.}, \au{Barkley, D.} \& \au{Hof, B.}} \yr{2023}
  \at{Transition to turbulence in pipe flow}.  \jt{Annu.\ Rev.\ Fluid Mech.}
  \bvol{55},  \pg{575--602}.

\bibitem[Avila {\em et~al.\/}(2010)Avila, Willis \& Hof]{Avila2010}
{\sc \au{Avila, M.}, \au{Willis, A.~P.} \& \au{Hof, B.}} \yr{2010}  \at{On the
  transient nature of localized pipe flow turbulence}.  \jt{J.\,Fluid Mech.}
  \bvol{646},  \pg{127--136}.

\bibitem[Barkley {\em et~al.\/}(2015)Barkley, Song, Mukund, Lemoult, Avila \&
  Hof]{Barkley2015}
{\sc \au{Barkley, D.}, \au{Song, B.}, \au{Mukund, V.}, \au{Lemoult, G.},
  \au{Avila, M.} \& \au{Hof, B.}} \yr{2015}  \at{The rise of fully turbulent
  flow}.  \jt{Nature}  \bvol{526},  \pg{550--553}.

\bibitem[Gom\'e {\em et~al.\/}(2020)Gom\'e, Tuckerman \& Barkley]{Gome2020}
{\sc \au{Gom\'e, S.}, \au{Tuckerman, L.~S.} \& \au{Barkley, D.}} \yr{2020}
  \at{Statistical transition to turbulence in plane channel flow}.  \jt{Phys.
  Rev. Fluids}  \bvol{5},  \pg{083905}.

\bibitem[Gom\'e {\em et~al.\/}(2022)Gom\'e, Tuckerman \& Barkley]{Gome2022}
{\sc \au{Gom\'e, S.}, \au{Tuckerman, L.~S.} \& \au{Barkley, D.}} \yr{2022}
  \at{Extreme events in transitional turbulence}.  \jt{Trans. R. Soc. A.}
  \bvol{380},  \pg{20210036}.

\bibitem[Hinrichsen(2000)]{Hinrichsen2010}
{\sc \au{Hinrichsen, H.}} \yr{2000}  \at{Non-equilibrium critical phenomena and
  phase transitions into absorbing states}.  \jt{Advances in Physics}
  \bvol{49:7},  \pg{815--958}.

\bibitem[Hof(2022)]{Hof2022}
{\sc \au{Hof, B.}} \yr{2022}  \at{Directed percolation and the transition to
  turbulence}.  \jt{Nature Reviews Physics}  \bvol{5},  \pg{62--72}.

\bibitem[Hof {\em et~al.\/}(2008)Hof, de~Lozar, Kuik \& Westerweel]{Hof2008}
{\sc \au{Hof, B.}, \au{de~Lozar, A.}, \au{Kuik, D.~J.} \& \au{Westerweel, J.}}
  \yr{2008}  \at{Repeller or attractor? selecting the dynamical model for the
  onset of turbulence in pipe flow.}  \jt{Phys.\ Rev.\ Lett.}  \bvol{101},
  \pg{214501}.

\bibitem[Hof {\em et~al.\/}(2006)Hof, Westerweel, Schneider \&
  Eckhardt]{Hof2006}
{\sc \au{Hof, B.}, \au{Westerweel, J.}, \au{Schneider, T.~M.} \& \au{Eckhardt,
  B.}} \yr{2006}  \at{Finite lifetime of turbulence in shear flows}.
  \jt{Nature}  \bvol{443},  \pg{59--62}.

\bibitem[Hugues \& Randriamampianina(1998)]{Hugues1998}
{\sc \au{Hugues, S.} \& \au{Randriamampianina, A.}} \yr{1998}  \at{An improved
  projection scheme applied to pseudospectral methods for the incompressible
  {N}avier-{S}tokes equations}.  \jt{Int. J.\,Num.\ Meth.\ Fluids}  \bvol{28},
  \pg{501--521}.

\bibitem[Kanazawa(2018)]{Kanazawa2018}
{\sc \au{Kanazawa, T.}} \yr{2018}  \at{Lifetime and growing process of
  localized turbulence in plane channel flow}. PhD thesis, Osaka University.

\bibitem[Kashyap {\em et~al.\/}(2020)Kashyap, Duguet \& Dauchot]{Kashyap2020}
{\sc \au{Kashyap, P.~V.}, \au{Duguet, Y.} \& \au{Dauchot, O.}} \yr{2020}
  \at{Flow statistics in the transitional regime of plane channel flow}.
  \jt{Entropy}  \bvol{22},  \pg{1001}.

\bibitem[Klotz {\em et~al.\/}(2022)Klotz, Lemoult, Avila \& Hof]{Klotz2022}
{\sc \au{Klotz, L.}, \au{Lemoult, G.}, \au{Avila, K.} \& \au{Hof, B.}}
  \yr{2022}  \at{Phase transition to turbulence in spatially extended shear
  flows}.  \jt{Phys.\ Rev.\ Lett.}  \bvol{128},  \pg{014502}.

\bibitem[Kohyama {\em et~al.\/}(2022)Kohyama, Sano \& Tsukahara]{Kohyama2022}
{\sc \au{Kohyama, K.}, \au{Sano, M.} \& \au{Tsukahara, T.}} \yr{2022}
  \at{Sidewall effect on turbulent band in subcritical transition of
  high-aspect-ratio duct flow}.  \jt{Phys.\ Fluids}  \bvol{34},  \pg{084112}.

\bibitem[Lawless(2003)]{Lawless2003}
{\sc \au{Lawless, Jerald~F.}} \yr{2003} {\em Statistical Models and Methods for
  Lifetime Data\/}, 2nd edn.  \publ{Hoboken, N.J.: Wiley-Interscience}.

\bibitem[Lemoult {\em et~al.\/}(2024)Lemoult, Mukund, Shih, Linga, Mathiesen,
  Goldenfeld \& Hof]{Lemoult2024}
{\sc \au{Lemoult, G.}, \au{Mukund, V.}, \au{Shih, H.-Y.}, \au{Linga, G.},
  \au{Mathiesen, J.}, \au{Goldenfeld, N.} \& \au{Hof, B.}} \yr{2024}
  \at{Directed percolation and puff jamming near the transition to pipe
  turbulence}.  \jt{Nat. Phys.}  \bvol{20},  \pg{1339--1345}.

\bibitem[Lemoult {\em et~al.\/}(2016)Lemoult, Shi, Avila, Jalikop, Avila \&
  Hof]{Lemoult2016}
{\sc \au{Lemoult, G.}, \au{Shi, L.}, \au{Avila, K.}, \au{Jalikop, S.~V.},
  \au{Avila, M.} \& \au{Hof, B.}} \yr{2016}  \at{Directed percolation phase
  transition to sustained turbulence in {C}ouette flow}.  \jt{Nat. Phys.}
  \bvol{12},  \pg{254--258}.

\bibitem[Manneville \& Shimizu(2020)]{Manneville2020}
{\sc \au{Manneville, P.} \& \au{Shimizu, M.}} \yr{2020}  \at{Transitional
  channel flow: A minimal stochastic model}.  \jt{Entropy}  \bvol{22},
  \pg{1348}.

\bibitem[Marensi {\em et~al.\/}(2023)Marensi, Yalniz \& Hof]{Marensi2023}
{\sc \au{Marensi, E.}, \au{Yalniz, G.} \& \au{Hof, B.}} \yr{2023}  \at{Dynamics
  and proliferation of turbulent stripes in plane-poiseuille and plane-couette
  flows}.  \jt{J.\,Fluid Mech.}  \bvol{974},  \pg{A21}.

\bibitem[Mukund {\em et~al.\/}(2025)Mukund, Paranjape, Sitte, Yalniz \&
  Hof]{Mukund2025}
{\sc \au{Mukund, V.}, \au{Paranjape, C.}, \au{Sitte, M.~P.}, \au{Yalniz, G.} \&
  \au{Hof, B.}} \yr{2025}  \at{Aging and memory of transitional turbulence}.
  \jt{Nature Communications}  \bvol{16},  \pg{8447}.

\bibitem[Paranjape(2019)]{Paranjape2019}
{\sc \au{Paranjape, C.~S.}} \yr{2019}  \at{Onset of turbulence in plane
  {Poiseuille} flow}. PhD thesis, IST Austria.

\bibitem[Peixinho \& Mullin(2006)]{Peixinho2006}
{\sc \au{Peixinho, J.} \& \au{Mullin, T.}} \yr{2006}  \at{Decay of turbulence
  in pipe flow}.  \jt{Phys.\ Rev.\ Lett.}  \bvol{96},  \pg{094501}.

\bibitem[Sano \& Tamai(2016)]{Sano2016}
{\sc \au{Sano, M.} \& \au{Tamai, K.}} \yr{2016}  \at{A universal transition to
  turbulence in channel flow}.  \jt{Nat. Phys.}  \bvol{12},  \pg{249--253}.

\bibitem[Shimizu \& Manneville(2019)]{Shimizu2019}
{\sc \au{Shimizu, M.} \& \au{Manneville, P.}} \yr{2019}  \at{Bifurcations to
  turbulence in transitional channel flow}.  \jt{Phys. Rev. Fluids}  \bvol{4},
  \pg{113903}.

\bibitem[Song \& Xiao(2020)]{Song2020}
{\sc \au{Song, B.} \& \au{Xiao, X.}} \yr{2020}  \at{Trigger turbulent bands
  directly at low {R}eynolds numbers in channel flow using a moving-force
  technique}.  \jt{J.\,Fluid Mech.}  \bvol{903},  \pg{A43}.

\bibitem[Tao {\em et~al.\/}(2018)Tao, Eckhardt \& Xiong]{Tao2018}
{\sc \au{Tao, J.~J.}, \au{Eckhardt, B.} \& \au{Xiong, X.~M.}} \yr{2018}
  \at{Extended localized structures and the onset of turbulence in channel
  flow}.  \jt{Phys. Rev. Fluids}  \bvol{3},  \pg{011902(R)}.

\bibitem[Tsukahara {\em et~al.\/}(2005)Tsukahara, Seki, Kawamura \&
  Tochio]{Tsukahara2005}
{\sc \au{Tsukahara, T.}, \au{Seki, Y.}, \au{Kawamura, H.} \& \au{Tochio, D.}}
  \yr{2005} {DNS} of turbulent channel flow at very low {R}eynolds numbers.
  \bt{In {\em Proceedings of Fourth International Symposium on Turbulence and
  Shear Flow Phenomena\/}},  \pg{pp. 935--940}. Williamsburg, USA.

\bibitem[Tuckerman {\em et~al.\/}(2014)Tuckerman, Kreilos, Shrobsdorff,
  Schneider \& Gibson]{Tuckerman2014}
{\sc \au{Tuckerman, L.~S.}, \au{Kreilos, T.}, \au{Shrobsdorff, H.},
  \au{Schneider, T.~M.} \& \au{Gibson, J.~F.}} \yr{2014}
  \at{Turbulent{-}laminar patterns in plane {P}oiseuille flow}.  \jt{Phys.\
  Fluids}  \bvol{26},  \pg{114103}.

\bibitem[Willis \& Kerswell(2007)]{Willis2007}
{\sc \au{Willis, A.~P.} \& \au{Kerswell, R.~R.}} \yr{2007}  \at{Critical
  behavior in the relaminarization of localized turbulence in pipe flow}.
  \jt{J.\,Fluid Mech.}  \bvol{98},  \pg{014501}.

\bibitem[Wu \& Song(2022)]{Wu2022}
{\sc \au{Wu, H.} \& \au{Song, B.}} \yr{2022}  \at{A numerical study of the
  side-wall effects on turbulent bands in channel flow at transitional
  {Reynolds} numbers}.  \jt{Computers \& Fluids}  \bvol{240},  \pg{105420}.

\bibitem[Wu \& Song(2025)]{Wu2025}
{\sc \au{Wu, Y.} \& \au{Song, B.}} \yr{2025}  \at{Transient characteristics of
  fully localized turbulence in transitional channel flow}.  \jt{J.\,Fluid
  Mech.}  \bvol{1009},  \pg{R2}.

\bibitem[Xiao \& Song(2020{\natexlab{{\em a\/}}})]{Xiao2020}
{\sc \au{Xiao, X.} \& \au{Song, B.}} \yr{2020{\natexlab{{\em a\/}}}}  \at{The
  growth mechanism of turbulent bands in channel flow at low {R}eynolds
  numbers}.  \jt{J.\,Fluid Mech.}  \bvol{883},  \pg{R1}.

\bibitem[Xiao \& Song(2020{\natexlab{{\em b\/}}})]{Xiao2020b}
{\sc \au{Xiao, X.} \& \au{Song, B.}} \yr{2020{\natexlab{{\em b\/}}}}
  \at{Kinematics and dynamics of turbulent bands at low {R}eynolds numbers in
  channel flow}.  \jt{Entropy}  \bvol{22},  \pg{1167}.

\bibitem[Xiong {\em et~al.\/}(2015)Xiong, Tao, Chen \& Brandt]{Xiong2015}
{\sc \au{Xiong, X.~M.}, \au{Tao, J.}, \au{Chen, S.} \& \au{Brandt, L.}}
  \yr{2015}  \at{Turbulent bands in plane-{P}oiseuille flow at moderate
  {R}eynolds numbers}.  \jt{Phys.\ Fluids}  \bvol{27},  \pg{041702}.

\bibitem[Xu \& Song(2022)]{Xu2022}
{\sc \au{Xu, D.} \& \au{Song, B.}} \yr{2022}  \at{Size-dependent transient
  nature of localized turbulence in transitional channel flow}.  \jt{J.\,Fluid
  Mech.}  \bvol{950},  \pg{R3}.

\bibitem[Zhang \& Tao(2023)]{Zhang2023}
{\sc \au{Zhang, L.} \& \au{Tao, J.}} \yr{2023}  \at{Large-scale mean flow and
  inclination of isolated turbulent band in channel flow}.
  \jt{arXiv:2310.03238} .

\end{thebibliography}

\end{document}